\documentclass[review]{elsarticle}

\usepackage[english]{babel}
\usepackage[T1]{fontenc}
\usepackage{etoolbox}
\makeatletter
\patchcmd{\ps@pprintTitle}{\footnotesize\itshape
      Preprint submitted to \ifx\@journal\@empty Elsevier
      \else\@journal\fi\hfill\today}{\scriptsize{Preprint submitted to Solar Energy \hfill \today}}{}{}

\def\ps@pprintTitle{%
   \let\@oddhead\@empty
   \let\@evenhead\@empty
   \let\@oddfoot\@empty
   \let\@evenfoot\@oddfoot
}

\makeatother
\usepackage{comment}
\usepackage{subfig}
\usepackage{hyperref}
\usepackage{amsmath}

\usepackage{amsmath,amssymb,amsfonts}
\usepackage{algorithmic}
\usepackage[]{graphicx}
\graphicspath{{Figures/}}
\usepackage{lipsum}
\usepackage{balance}
\usepackage{textcomp}
\usepackage{multirow}
\usepackage{color}
\usepackage{lscape}
\usepackage{longtable,booktabs}
\usepackage{seqsplit}
\begin{document}

\setcounter{page}{0} 

\begin{frontmatter}
\title{Air Quality in the New Delhi Metropolis under COVID-19 Lockdown} 

\author[add1,add2]{Dewansh~Kaloni\corref{details}}\cortext[details]{A part of this research was conducted while D. Kaloni was doing an internship at ADAPT SFI Research Centre, Dublin, Ireland.}
\ead{f20170688@pilani.bits-pilani.ac.in}
\author[add3]{Yee~Hui~Lee}
\ead{EYHLee@ntu.edu.sg}
\author[add2,add4]{Soumyabrata~Dev\corref{mycorrespondingauthor}}
\cortext[mycorrespondingauthor]{Corresponding author. Email: soumyabrata.dev@ucd.ie,  Tel.: + 353 1896 1797.}
\ead{soumyabrata.dev@ucd.ie}
\address[add1]{Birla Institute of Technology and Science, Pilani, India}
\address[add2]{ADAPT SFI Research Centre, Dublin, Ireland}
\address[add3]{School of Electrical and Electronic Engineering, Nanyang Technological University, Singapore, Singapore}
\address[add4]{School of Computer Science, University College Dublin, Ireland}

\begin{abstract}
Air pollution has been on continuous rise with increase in industrialization in metropolitan cities of the world. Several measures including strict climate laws and reduction in the number of vehicles were implemented by several nations. The COVID-19 pandemic provided a great opportunity to understand the daily human activities effect on air pollution. Majority nations restricted industrial activities and vehicular traffic to a large extent as a measure to restrict COVID-19 spread. In this paper, we analyzed the impact of such COVID19-induced lockdown on the air quality of the city of New Delhi, India. We analyzed the average concentration of common gaseous pollutants \textit{viz.} sulfur dioxide (SO$_2$), ozone (O$_3$), nitrogen dioxide (NO$_2$), and carbon monoxide (CO). These concentrations were obtained from the tropospheric column of Sentinel-5P (an earth observation satellite of European Space Agency) data. We observed that the city observed a significant drop in the level of atmospheric pollutant's concentration for all the major pollutants as a result of strict lockdown measure. Such findings are also validated with pollutant data obtained from ground based monitoring stations. We observed that near-surface pollutant concentration dropped significantly by 50\% for PM$_{2.5}$, 71.9\% for NO$_2$, and 88\% for CO, after the lockdown period. Such studies would pave the path for implementing future air pollution control measures by environmentalists. 
\end{abstract}


\end{frontmatter}

\section{Introduction}

\label{sec:intro}
Air pollution is one of the most important sustainability
concerns faced by most of the developed nations of the
world~\cite{agrawal2021investigation}. It is identified in the
Sustainable Development Goals (SDG): Goal 3 (Good Health and
Well-being)~\cite{guegana2018sustainable} and Goal 11
(Sustainable Cities and Communities)~\cite{vaidya2020sdg}.
Gu\'egana \textit{et al.}  in~\cite{guegana2018sustainable} discusses SDG:
Goal 3 (Good Health and Well-being) and the need for integrative
thinking in public and animal health for developing countries.
Furthermore, Vaidya and Chatterji in~\cite{vaidya2020sdg} discuss
SDG 11 and the new urban agenda for addressing global
sustainability at an urban scale. The occurrences of forest fires
that happen at sporadic intervals also lead to increased air
pollution~\cite{singh2021parallel}. The recent onset of COVID-19
(C19) and its associated lockdown led to an increased awareness
of this pressing issue.  Jia \textit{et al.}  in
~\cite{jia2020insignificant} showed insignificant impact on air
quality due to lockdown in Memphis, US using ground based
monitoring data, in comparison to previous months. Similar
observations were observed in the city of New York, US, where
Zangari \textit{et al.}  observed that there was no significant air quality
changes in New York because of the COVID-19
pandemic~\cite{zangari2020air}. On the other hand, a city like
Delhi, India known for severe air pollution issues showed a large
contrast for noticeable observations in our study both with
satellite and ground based data.  

The lockdown induced by the COVID-19 pandemic observed a decrease
in atmospheric pollutants around the globe. Owing to the complete
halt of vehicular traffic and reduction in industrial activities,
we observed a positive impact on the amount of atmospheric
pollutants and the overall air quality. The lockdown because of
COVID-19 pandemic has provided the environmental researchers an
opportunity to analyze the atmospheric pollutants in finer scale.
Bourdrel \textit{et al.}  in~\cite{bourdrel2021impact} attempted to provide
a relationship between air quality and chronic diseases in the
context of COVID-19. Similar work on the relationship between air
pollution and COVID-19 infections and mortality was performed by
Ali and Islam in~\cite{ali2020effects}. Furthermore, Nigam \textit{et al.}
in~\cite{nigam2021positive} identified the positive effects of
COVID-19 in the industrial cities of Ankleshwar and Vapi in
India. They noted a massive decrease in the concentration of
NO$_2$ in the atmosphere, owing to the halt in industrial
activities. Karaer in~\cite{karaer2020analyzing} analyzed the 67
Florida Counties to present the strong correlation between the
drop in NO2 concentrations (from Sentinel-5p) and Vehicle Mile
Traveled (VMT) estimates from cell phone mobility records. Kaloni
\textit{et al.}  performed a systematic analysis on the impact of COVID-19
on the air pollution of Dublin city~\cite{kaloni2021impact}.
Franch-Pardo \textit{et al.}  in~\cite{franch2021review} provided a
systematic review of 221 scientific articles published in 2020
that used spatial and GIS technologies to understand the dynamics
of COVID-19. Doolette \textit{et al.}  investigated the atmosphere from the
context of nanoparticle fertilizer in precision
agriculture~\cite{doolette2020zinc}. Such recent works reveal
that there is a promising opportunity for researchers to analyze
the atmospheric pollutants, in the context of COVID-19 pandemic.

In this paper, we focused our case study on the city of New Delhi
in India. The primary reasons of choosing New Delhi was because
this city has a historical problem of poor air quality, and
therefore it provided us an opportune moment in performing such
atmospheric analysis. The city of New Delhi has its long-lasting
history with high pollution levels~\cite{alparslan2021analyzing}.
Traditionally, government introduced several measures including
plying of odd and even numbered vehicles on alternative days, and
introduction of CNG vehicles~\cite{cngreport,mathur2019impact}.
However, there was an insignificant improvement in Air Quality
Index (AQI)~\cite{mohan2007analysis}. The AQI level stayed in
\textit{severe} and \textit{very poor} category~\cite{explo}.
However, with the introduction of lockdown owing to C19 pandemic,
we observed a significant improvement in AQI level, and reached
the \textit{moderate} category during the lockdown .

In this paper,\footnote{In the spirit of reproducible research, code used to obtain plots and results is shared at  \url{https://github.com/dkaloni/LockdownAnalysis}.} we discussed the impact of the lockdown induced by the C19 pandemic on the atmospheric air quality. The novel contributions of this paper are as follows:
\begin{itemize}
    \item Systematic analysis of the impact of lockdown on air quality;
    \item Subjective and objective validation of improved air quality from satellite and ground-based data; and
    \item Release of open-source code and datasets for further benchmarking by the community.
\end{itemize}

The rest of the paper is arranged as follows.  Section~\ref{sec:dataset} discusses the TROPOMI instrument of Sentinel-5P data that is used for our analysis and the data format of files obtained from the sources.  Section~\ref{sec:approach} discusses with the methodology used to obtain large sized satellite data and resolving it to usable form.  Section~\ref{sec:satellite} briefly explains the change in nitrogen dioxide over the Central India region, followed by discussion on concentration of harmful gases in the atmosphere of Delhi using satellite data.   Section~\ref{sec:ground} correlate satellite data with the ground-based monitoring station data and thereby confirming our findings with regard to harmful gases. Finally,  Section~\ref{sec:conc} concludes the paper and discusses the future work.

\section{Air Quality Monitoring Datasets}

\label{sec:dataset}
Copernicus, an European Union's Earth Observation Programme consists of several satellites. The Sentinel-5P is one of the Earth Orbit Polar Satellite System collecting data on trace gases concentration, climate and air quality.

\subsection{TROPOMI}

TROPOMI stands for TROPOspheric Monitoring Instrument and is a
satellite instrument attached to Sentinel-5P to examine the air
quality with the best accuracy to
date~\cite{guanter2015potential}. TROPOMI consists of four
detectors that can detect wavelength in the shortwave infrared,
near-infrared, and UV--visible covering a large portion of
spectrum range. TROPOMI provides the best resolution data to
study concentration of gases like sulfur dioxide, nitrogen
dioxide, formaldehyde, methane, ozone, carbon monoxide, and
aerosols present in the atmosphere~\cite{tropomidetails}.

Sentinel-5P's TROPOMI is the primary source of data for this study. It consists of two-dimensional detector collecting data in strips of Earth. Satellite moves $7$ km in the period of nearly $1$s. Direction along the track of the satellite is $2600$ km whereas along the track direction it is $7$ km.

 \subsection{Ground-Based Monitoring Station Data}

In addition to the satellite
data~\cite{sun2015slope,das2021estimating}, ground-based
observation data can perfectly complement the satellite data in
monitoring the atmospheric events. These ground-based observation
data include weather stations~\cite{akrami2021graph}, rain
gauges~\cite{manandhar2019data}, pyranometers and ground-based
sky imagers~\cite{dev2019estimating,dev2017color}. These
ground-based observations provided us a fantastic alternative
approach on the study of air quality~\cite{danesi2021monitoring,danesi2021predicting}.
Such observations provided us a near-accurate estimate of the
level of atmospheric pollutants on the surface of the earth.
Therefore, we obtained a complete analysis on the effect of
lockdown on the level of atmospheric pollutants --- from both
satellite and ground-based observation. In our study, the
ground-based data was collected from ITO station of Central
Pollution Control Board (CPCB).\footnote{The source of the
ground-based dataset is
\url{https://app.cpcbccr.com/ccr/}.}

\begin{figure}[htb]
\centering
\subfloat[19-Mar-2020 \newline(before lockdown-I)]{\includegraphics[height=0.2\textwidth]{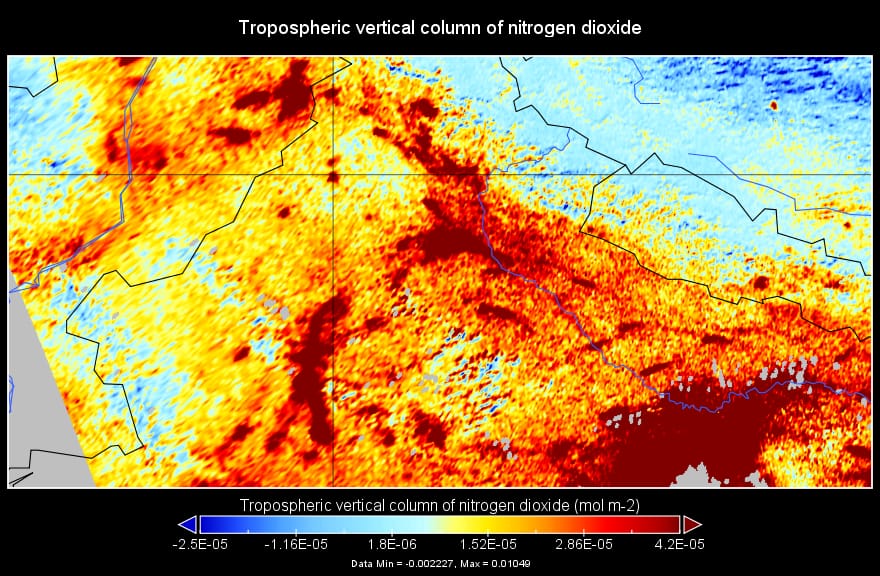}}
\subfloat[22-Mar-2020 \newline(on lockdown-I)]{\includegraphics[height=0.2\textwidth]{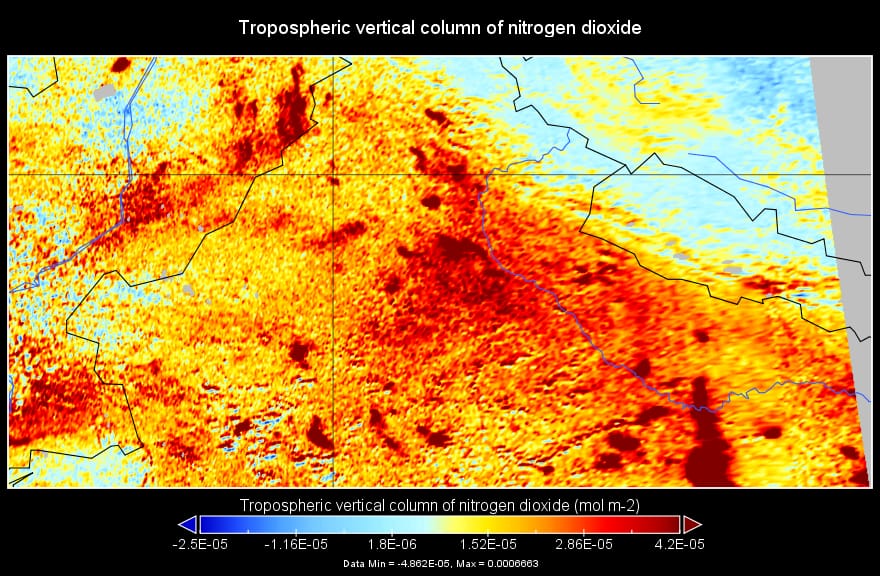}}
\subfloat[23-Mar-2020 \newline(lockdown-I lifted)]{\includegraphics[height=0.2\textwidth]{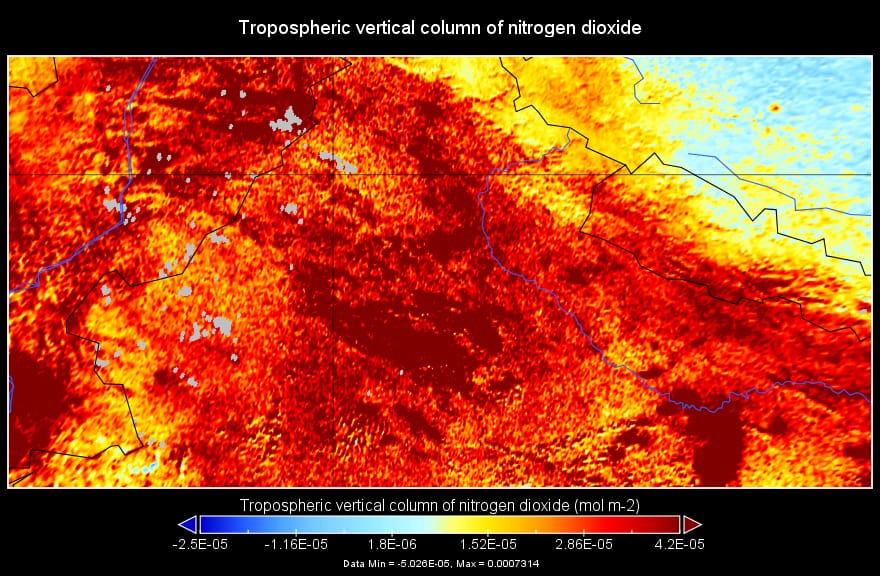}}\\
\subfloat[24-Mar-2020]{\includegraphics[height=0.2\textwidth]{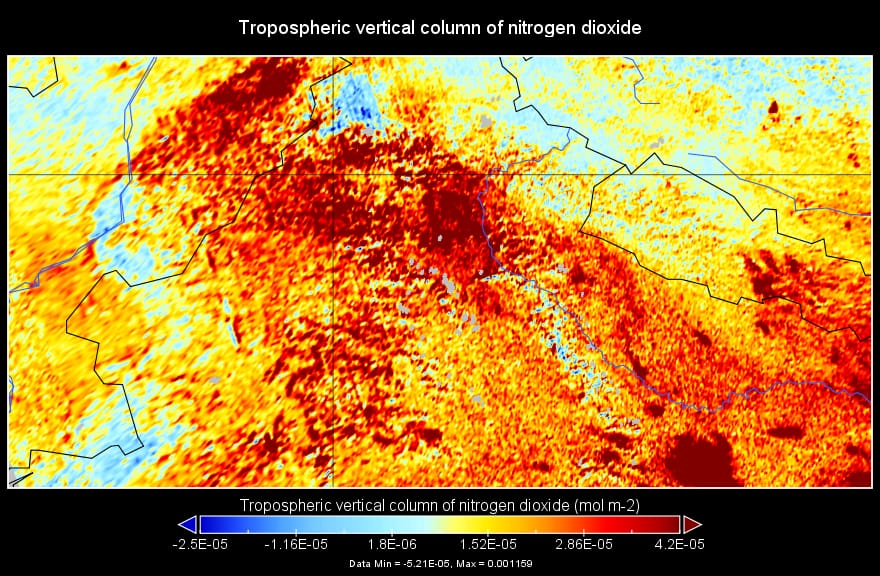}}
\subfloat[28-Mar-2020 \newline(during lockdown-II)]{\includegraphics[height=0.2\textwidth]{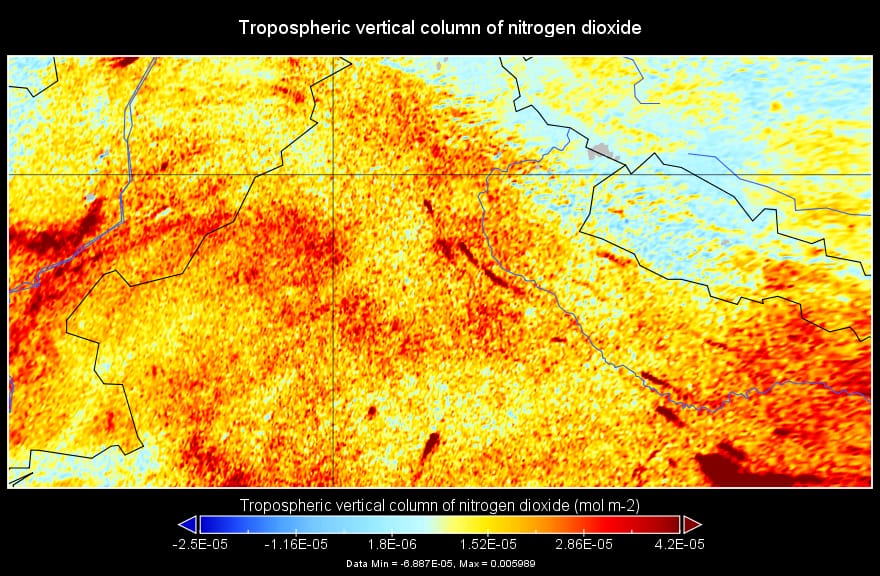}}
\subfloat[29-Mar-2020 \newline(during lockdown-II)]{\includegraphics[height=0.2\textwidth]{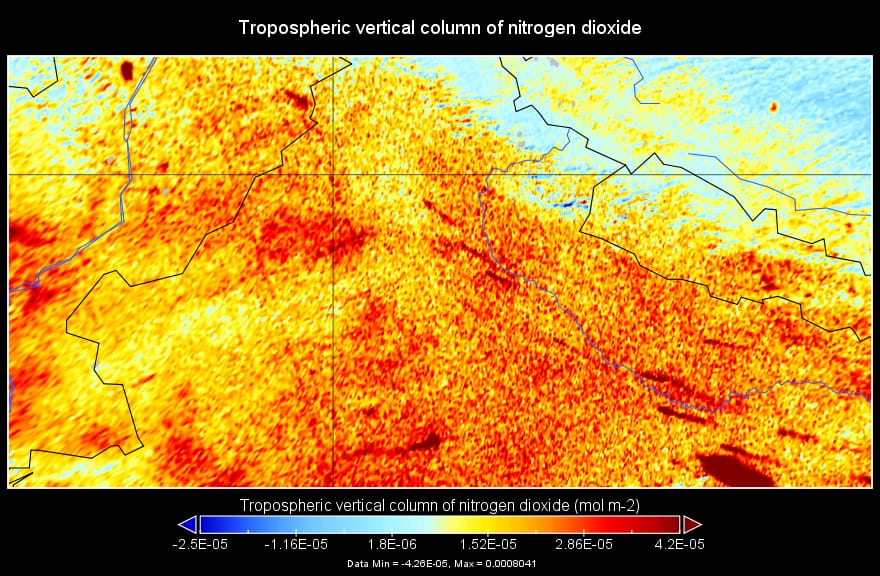}}
\caption{We observed that the concentration of NO$_2$ dropped significantly on 22-Mar-2020 (day of the lockdown-I) for the northern region of India. We noticed a sharp increase again on 23-Mar-2020 as it was a normal day where lockdown was lifted. The second lockdown was again imposed from 25-Mar-2020 for a period of 21 days and we observed a clear reduction in NO$_2$ concentration (\textit{cf.} e, f submaps).}
\label{fig:nindia-no2}
\end{figure}

India confirmed its first confirmed case of COVID-19 on 30-Jan-2020, and thereby saw a slow growth in the number of cases during the month of February. Cases count increased sharply in the middle of the month of March, and consequently  educational institutes were closed down to control the spread. Subsequently, a strict lockdown to further curtail the spread was imposed throughout the nation. The lockdown was implemented in stages. The lockdown-I was done for a single day on 22-Mar-2020 (popularly known as the \textit{janta curfew}). Subsequently, the upcoming lockdown-II was imposed on 25-Mar-2020 and remained enforced till 14-Apr-2020, to curtain the spread and avoid community transmission as much as possible.

This presented a surprisingly great reduction in pollution
levels, with AQI shifted to the \textit{moderate} category
clearly directed the effectiveness of lockdown with respect to
drop in air pollution.  Fig.~\ref{fig:nindia-no2} demonstrates that
with complete closure of factories and vehicle movement near to
zero, the environment saw a sudden drop of about 36.6\% in the
average nitrogen dioxide tropospheric column concentration in the
atmosphere above Delhi on 22-Mar-2020. This trend in the
reduction of the concentration of NO$_2$ continued for the
subsequent days as well~\cite{srivastava202021}. There was a second lockdown on 25-Mar-2020, and we observed a low concentration of NO$_2$ on 28-Mar-2020 and 29-Mar-2020, as illustrated in Fig.~\ref{fig:nindia-no2}(e) and Fig.~\ref{fig:nindia-no2}(f).

\section{Our Approach}

\label{sec:approach}

This study was mainly focused on Delhi and involved processing data specifically for the Delhi region. We systematically analyzed the data obtained from both satellite and ground-based observations. The data obtained from TROPOMI were stored in NetCDF format. The NetCDF stands for Network Common Data Form, and is the format of file used for multidimensional data to study atmospheric features. The Sentinel-5P data uses NetCDF files to share and store files in an array-like structure stacked together. It is generally used to share climate data and for related purposes.  The NetCDF4 file contains both the data and the metadata for the product.

\begin{figure*}
\centering
\subfloat[Nitrogen Dioxide ]{\includegraphics[height=0.35\textwidth]{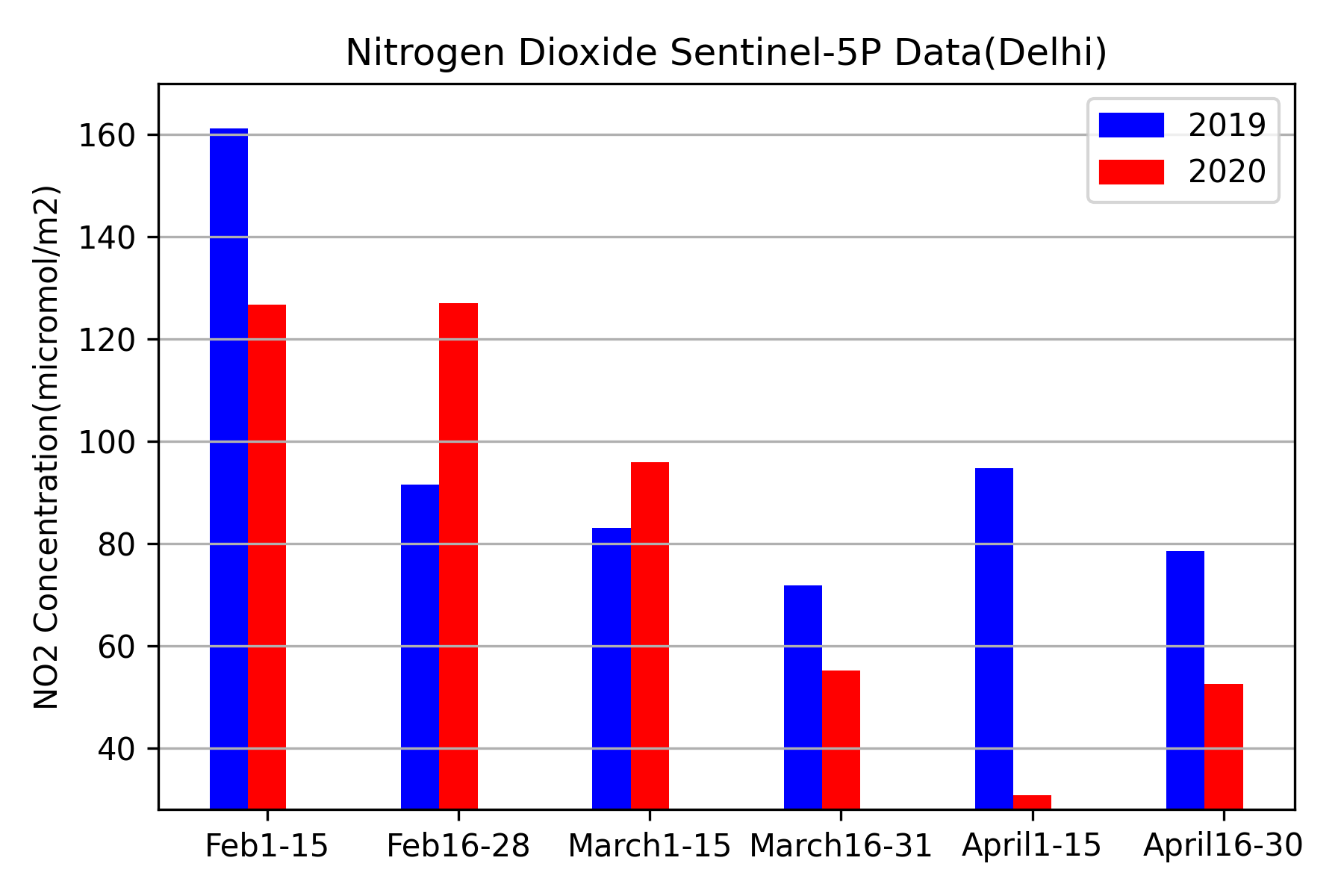}}
\subfloat[Sulphur Dioxide]{\includegraphics[height=0.35\textwidth]{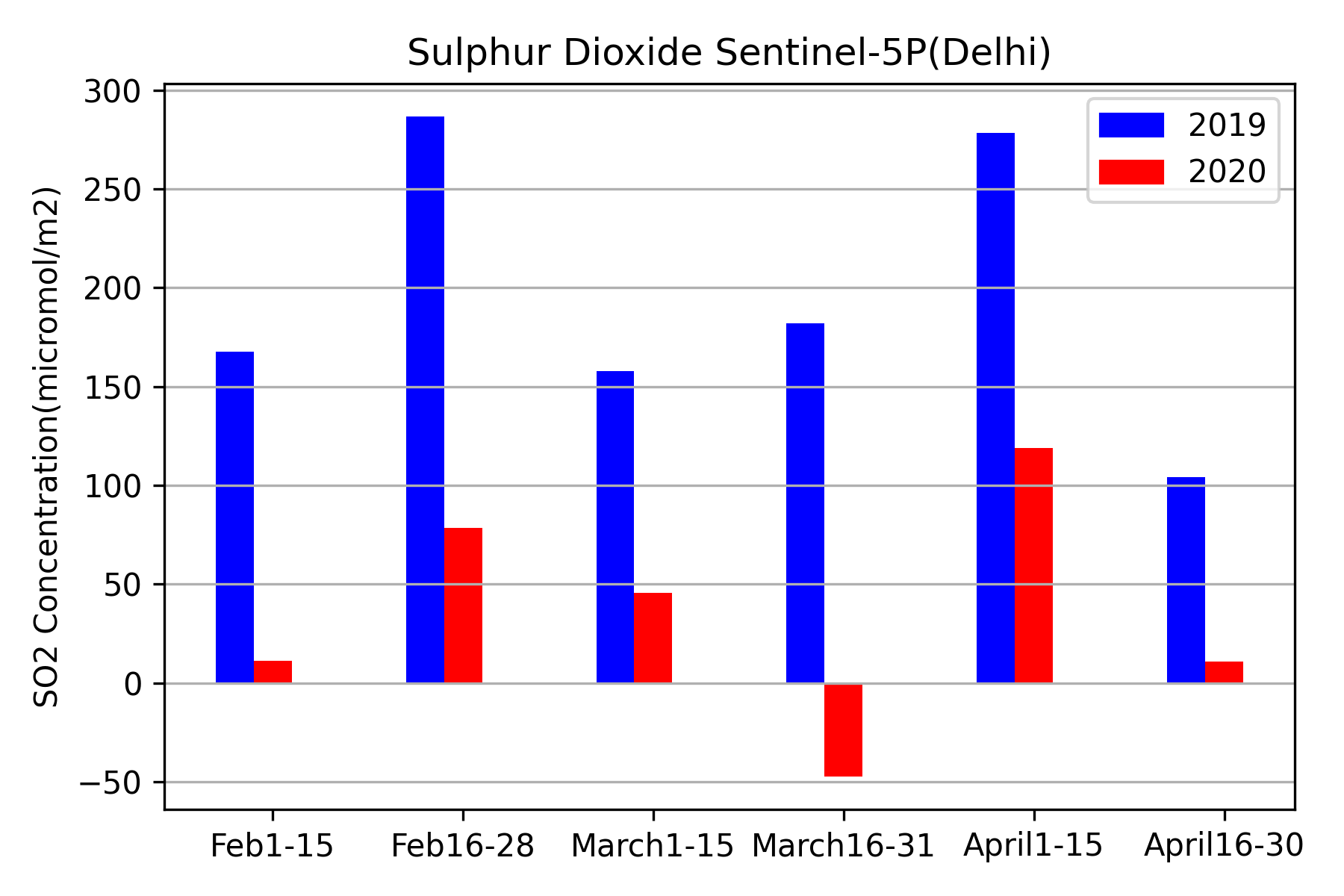}}\\
\subfloat[Carbon Monoxide]{\includegraphics[height=0.35\textwidth]{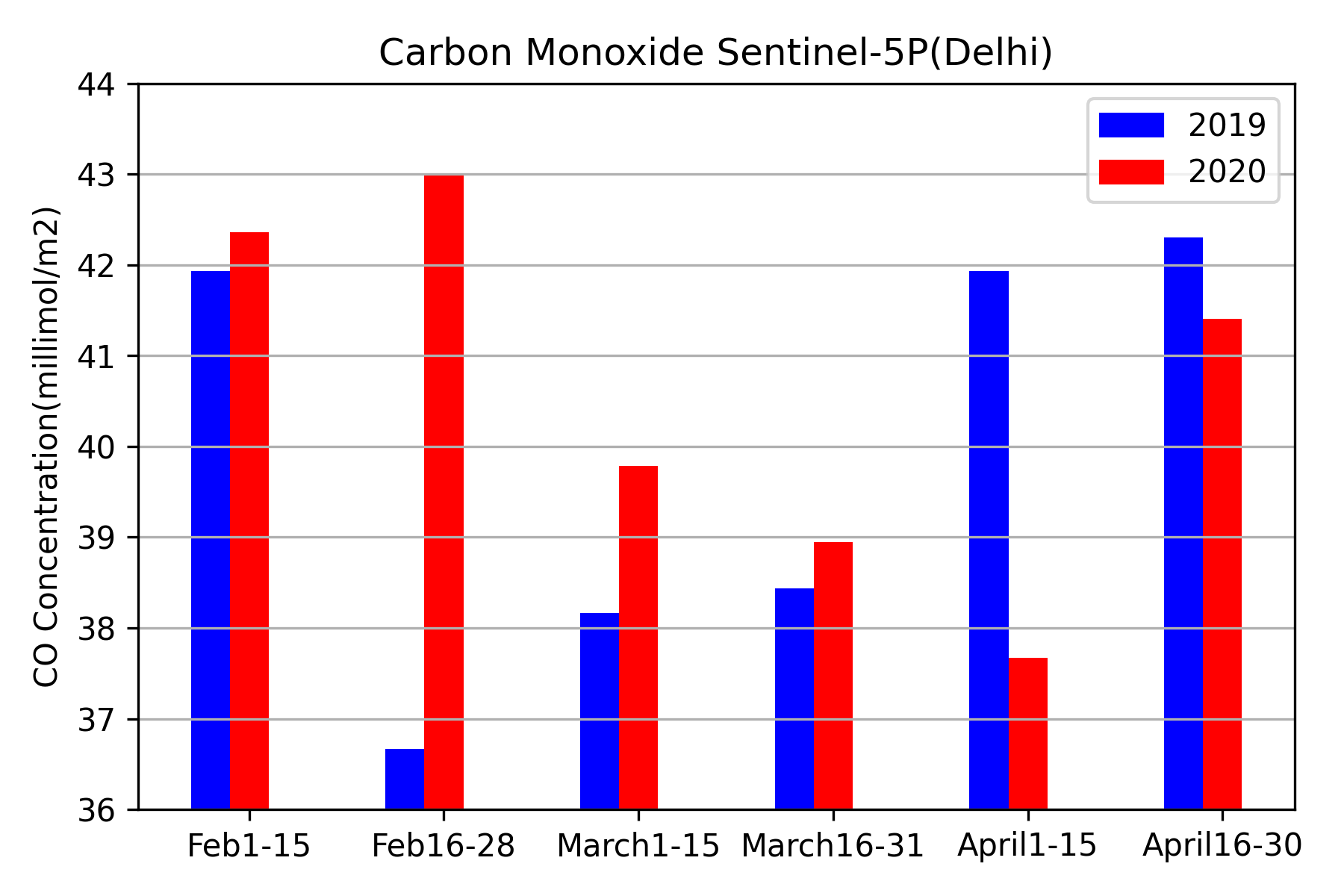}}
\subfloat[Ozone]{\includegraphics[height=0.35\textwidth]{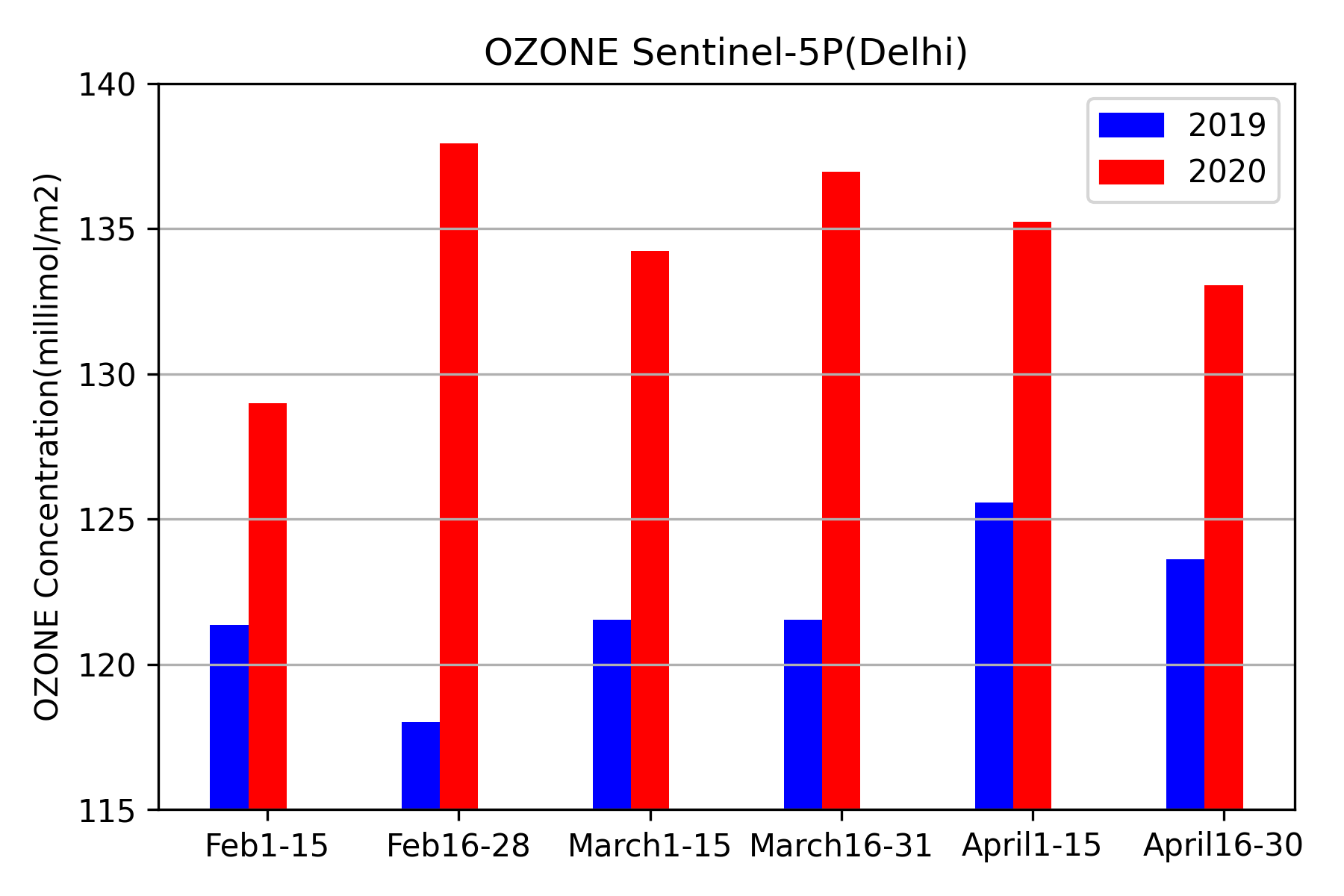}}
\caption{Distribution of various atmospheric pollutants obtained from TROPOMI satellite data across the lockdown period. We observed: (a) clear reduction in average concentration of NO$_2$ for the period of 16-Mar-2020 till 31-Mar-2020 in comparison to previous $15$ days.  
Such observation can also be observed with respect to the previous year of 2019, 
(b) SO$_2$ concentration saw a sharp drop in average concentration for the period of 16-Mar-2020 till 31-Mar-2020. In comparison to 2019, lockdown period caused a significant drop in SO$_2$ concentration, (c) the concentration of CO during the period of 1-Apr-2019 till 15-Apr-2019 with respect to the similar period of 1-Apr-2020 till 15-Apr-2020 recorded a significant drop of 10\%, (d) the concentration of O$_3$ in Apr-2020 increased by $7$\% as compared to that of Apr-2019~\cite{nagappa2020now}. }
\label{fig:satellite-analysis}
\end{figure*}

To subset out New Delhi from the global TROPOMI dataset, we performed a sequence of methodological steps in order to subset out the data of New Delhi from the global dataset. The selection of New Delhi region was performed by providing the range of latitude and longitude of the region. We subset out the Delhi region by providing the latitude range from $28.40^{\circ}$ to $28.90^{\circ}$ and longitude range from $76.80^{\circ}$ to $77.35^{\circ}$. The data points included in this region of interest were thereby used for the subsequent computation of pollutant concentration.  The measurements were stored and archived in individual NetCDF files. Each NetCDF file indicated the concentration of a trace gas for the selected day of the study.  We then combined all the obtained files into a single file by adding a new dimension of time to the dataset for comparative study.\footnote{The NCO is a set of command line utilities used to operate NetCDF files efficiently, available at \url{http://nco.sourceforge.net/}.} This combined file provided us with data points for the area spread across the time dimension.

Our obtained combined NetCDF file provided us the estimate of the average daily concentration of the various atmospheric pollutants. We processed these files to obtain the pollutant concentrations of the different pollutant gases. We replaced the \textit{NaN} values of the data grid to $0$, and thereby calculated the sum of the all the measurement values in the particular grid. We divided the resultant sum by the number of non-zero values to have a daily average for the region. This enabled us to calculate the average concentration of a particular pollutant over the selected region of New Delhi across a 15-days period. These steps were subsequently repeated for the different gases in the tropospheric column.

\section{Satellite-based Pollutants Analysis}

\label{sec:satellite}

In this section, we delved further into the concentration of the individual pollutants obtained from the TROPOMI dataset. We compared the average concentration of such pollutant with that of the previous year data to quantify the improvement in the air quality.

\subsection{Nitrogen Dioxide}

The data used for this study is
\texttt{nitrogendioxide\_tropospheric\_column} product of the
file which gives the total atmospheric NO$_2$ column
between the surface and the top of the troposphere, which stands
for tropospheric vertical column of nitrogen
dioxide~\cite{NO2Sent1}.

We performed the primary analysis on the subset of central India. We generate the spatial concentration maps of nitrogen dioxide (\textit{cf.}  Fig.~\ref{fig:nindia-no2})  using \texttt{panoply}, a geo-referenced data viewer by NASA Goddard Institute of Space Studies. The nitrogen dioxide concentration was above average on 19-Mar-2020, which dropped significantly on 22-Mar-2020 (day of \textit{janta curfew}). It noticed a sharp increase again on 23-Mar-2020 as it was a normal day, and vehicles, factories operated in full strength increasing NO$_2$ emission in the atmosphere. However, since lockdown was again imposed from 25-Mar-2020 for 21 days, the improvement can be clearly noticed on~Fig.~\ref{fig:nindia-no2}(e) and  Fig.~\ref{fig:nindia-no2}(f). We also noticed reduction in NO$_2$ concentration in the troposphere during the initial days of lockdown-II.

Fig.~\ref{fig:satellite-analysis} shows that Delhi observed a similar trend and drop in concentration was clear. In Delhi, the calculated average concentration of nitrogen dioxide was $95.9$ $\mu$mol/m$^2$ for first half of Mar-2020 and dropped to $55.2$ $\mu$mol/m$^2$ for the second half of Mar-2020. The condition further improved, and the average NO$_2$ concentration dropped to $30.8$ $\mu$mol/m$^2$ for the first fortnight of Apr-2020. The plots clearly show that concentration continued to drop after lockdown. However, with the ease of lockdown towards the end of Apr-2020, it again started increasing.

\subsection{Sulfur Dioxide}

The data used for this study is
\texttt{sulphurdioxide\_total\_vertical\_column} product of the
NetCDF file which provided the total atmospheric column between
the surface and the top of the troposphere~\cite{SO2Sent1}.
Fig.~\ref{fig:satellite-analysis} shows that the average
concentration for the first fortnight of Mar-2020 was
$45.6$ $\mu$mol/m$^2$ which dropped to $-47.4$
$\mu$mol/m$^2$. The negative vertical column values
are an indication of sensor detecting low concentration of the
pollutant.\footnote{Please refer to page 9, with the heading
`Negative Vertical Column Density values' from the product manual
available here:
\url{https://sentinels.copernicus.eu/documents/247904/3541451/Sentinel-5P-Sulphur-Dioxide-Readme.pdf}.}
As per the product manual~\cite{So2ProductManual} these values
should not be filtered except for outliers,  \textit{i.e.} for
vertical columns lower than $0.001$ $\mu$mol/m$^2$. This
clearly indicates a drop in SO$_2$ concentration to such a
low level that region was under clear category. But, this
increased to $119$ $\mu$mol/m$^2$ for the
first fortnight of Apr-2020 due to the other meteorological
factors and again dropped to $10.8$ $\mu$mol/m$^2$ for the
second fortnight of Apr-2020. These other meteorological factors
include surface wind speed, surface pressure, relative humidity
and 850-mb level temperature. These meteorological parameters
impact the pollutant concentration and introduce variability in
its concentration. In comparison to 2019 data for the same
period, the concentration of SO$_2$ was
$182.2$ $\mu$mol/m$^2$ for 16--31, Mar-2019,
$278.35$ $\mu$mol/m$^2$ for 1--15, Apr-2019 and
$103.98$ $\mu$mol/m$^2$ for 16--30, Apr-2019. The sulfur
dioxide concentration for the month of Apr-2020 saw a drop of
about 66\% in comparison to Apr-2019 average concentration.

\subsection{Carbon Monoxide}

The data used for analyzing carbon monoxide concentration was
\texttt{carbonmonoxide\_total\_column} product of the NetCDF file, which
provided the total atmospheric column between the surface and the
top of the atmosphere~\cite{COSent}.
Fig.~\ref{fig:satellite-analysis} shows that the Delhi's average
concentration for Mar-2019 was 38.79 mmol/m$^2$, whereas
in Mar-2020, the recorded concentration was 39.36
mmol/m$^2$. The first fortnight of Apr-2019 recorded
concentration of 41.92 mmol/m$^2$, and the first fortnight
of Apr-2020 recorded 37.67 mmol/m$^2$. This indicated a
drop of 10\% indicating that the lockdown showed significant
impact in the reduction of concentration of CO in the first half
of April.

\subsection{Ozone}

  Another harmful gas is ozone which is not directly emitted but
is a product of reactions between volatile organic compounds and
oxides of nitrogen (NO$_x$) in presence of sunlight.
Previous several years of data showed that ozone concentration
usually increases during spring and summer season in National
Capital Region of India~\cite{airqual2020}. The data used for
analyzing Ozone concentration was
\texttt{ozone\_total\_vertical\_column} product of the NetCDF
file  which provided the total atmospheric column between the
surface and the top of the atmosphere~\cite{OZONESent1}. We
observed from  Fig.~\ref{fig:satellite-analysis} that ozone
concentration for Apr-2019 was 124.59 mmol/m$^2$ whereas
that increased by 7\% to 134.14 mmol/m$^2$ for the month
of Apr-2020. Similar increase in concentration was observed for
the month of March.  One of the stated reasons for increased
levels of ozone concentration in the atmosphere of Delhi is that,
ozone forms when NO$_{x}$ and volatile gases react under the
influence of sunlight and temperature. But the cyclical chemistry
of ozone suggests that ozone again reacts with the NO$_{x}$
present in the atmosphere and gets removed. Hence, even though
ozone is formed in high NO$_{x}$ concentration areas with
other gases providing suitable conditions, it does not stay.
However, the ozone has a longer resident life and builds up in
its concentration, when it moves to less polluted areas
(\textit{i.e.} Delhi during lockdown).  The NO$_2$
concentration is considered a good representative of
NO$_{x}$.

\subsection{Impact of lockdown}

\begin{figure}[htb]
\begin{center}
\includegraphics[width=0.9\textwidth]{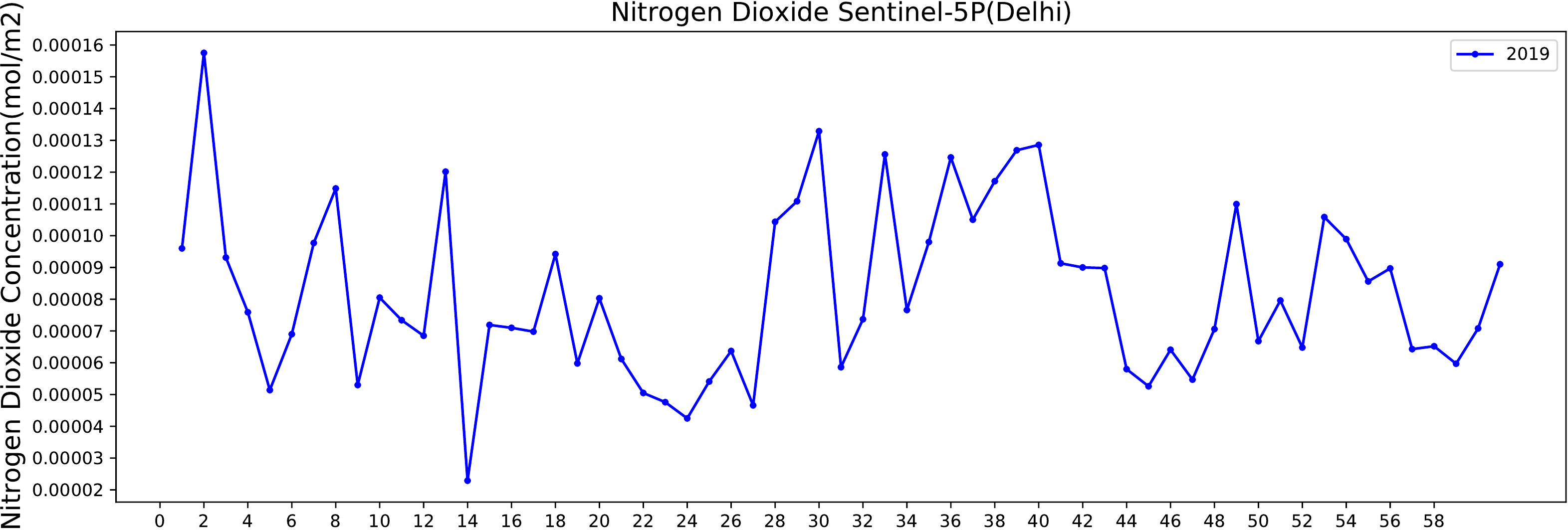}\\
\vspace{3mm}
\includegraphics[width=0.9\textwidth]{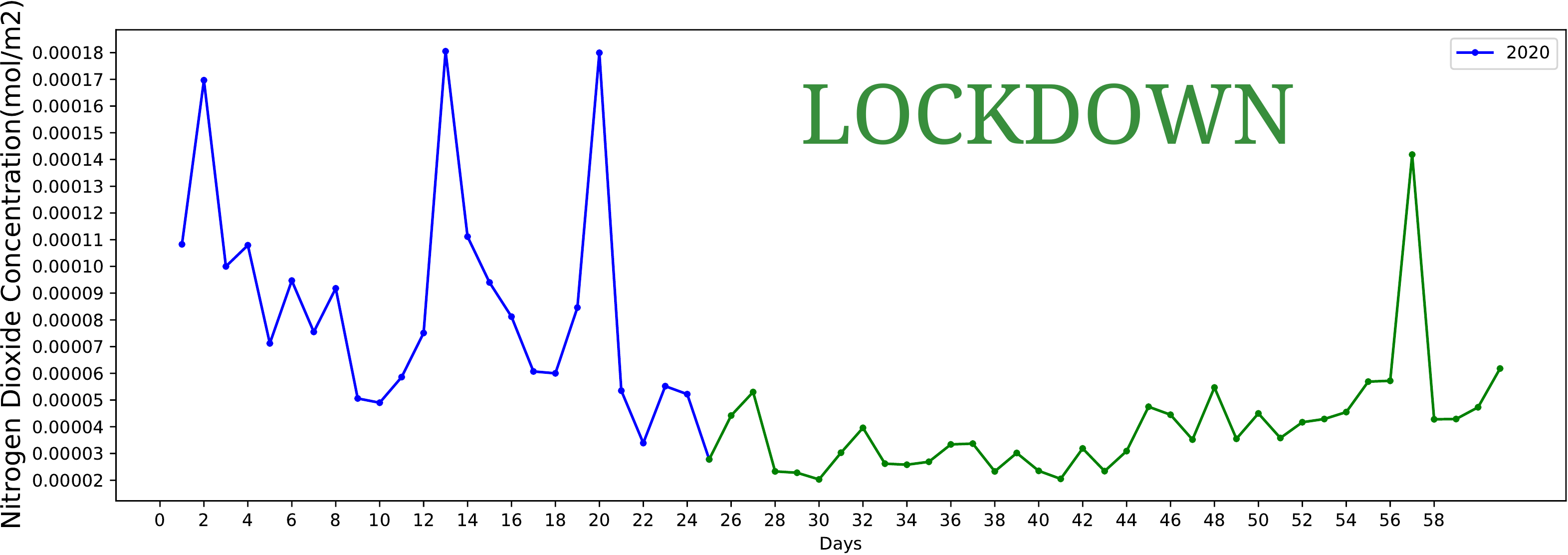}
\caption{Plot of average daily concentration of nitrogen dioxide obtained from Sentinel-5P for the years of 2019 and 2020. The x-axis indicates the days starting from 1-Mar-2019 (top graph) and 1-Mar-2020 (below graph). The y-axis indicates the nitrogen dioxide concentration. This time series clearly shows that the lockdown significantly reduced the concentration of nitrogen dioxide for the similar period of the year. 
\label{fig:lockdown-timeseries}}
\end{center}
\end{figure}

Finally, for comparative study for month of March and April between 2019 and 2020 we obtained daily average concentration of Nitrogen Dioxide for Delhi from the complete dataset and plot each day average concentration as data points on a time-series plot.  Fig.~\ref{fig:lockdown-timeseries} shows the impact of the lockdown on the concentration of nitrogen dioxide in the atmosphere. The $x$-axis in  Fig.~\ref{fig:lockdown-timeseries} indicates the days starting from 1-Mar-2019 (top graph) and 1-Mar-2020 (below graph) respectively. The $y$-axis indicates the daily concentration of nitrogen dioxide in the atmosphere. We observed similar trend in NO$_2$ concentration for the first half of March 2020 as compared to previous year. This is then followed by a sharp drop after 25-Mar-2020. The NO$_2$ concentration dropped significantly after lockdown was imposed. We indicated the days in lockdown in green color in  Fig.~\ref{fig:lockdown-timeseries} for clarity. Post lockdown, the nitrogen dioxide maintained a significant lower value as compared to that of the same period of 2019.

\section{Ground-based Pollutants Analysis}

\label{sec:ground}
The ground-based pollutant data was collected for ITO station of Central Pollution Control Board, Ministry of Environment, India. It is one of the largest atmosphere monitoring station in India. The data is freely accessible from {\url{https://app.cpcbccr.com/ccr}}. We collected and analyzed data for a period of two years for the years 2019 and 2020. We collected the Air Quality Index (AQI), Particulate Matter 10 (PM10), Particulate Matter 2.5 (PM2.5), and other pollutant gases present in the atmosphere. We specifically plotted the daily concentration data for the months of February, March and April across the years because the lockdown came into effect during these months.

\begin{figure*}
\centering
\subfloat[Nitrogen Dioxide ]{\includegraphics[height=0.35\textwidth]{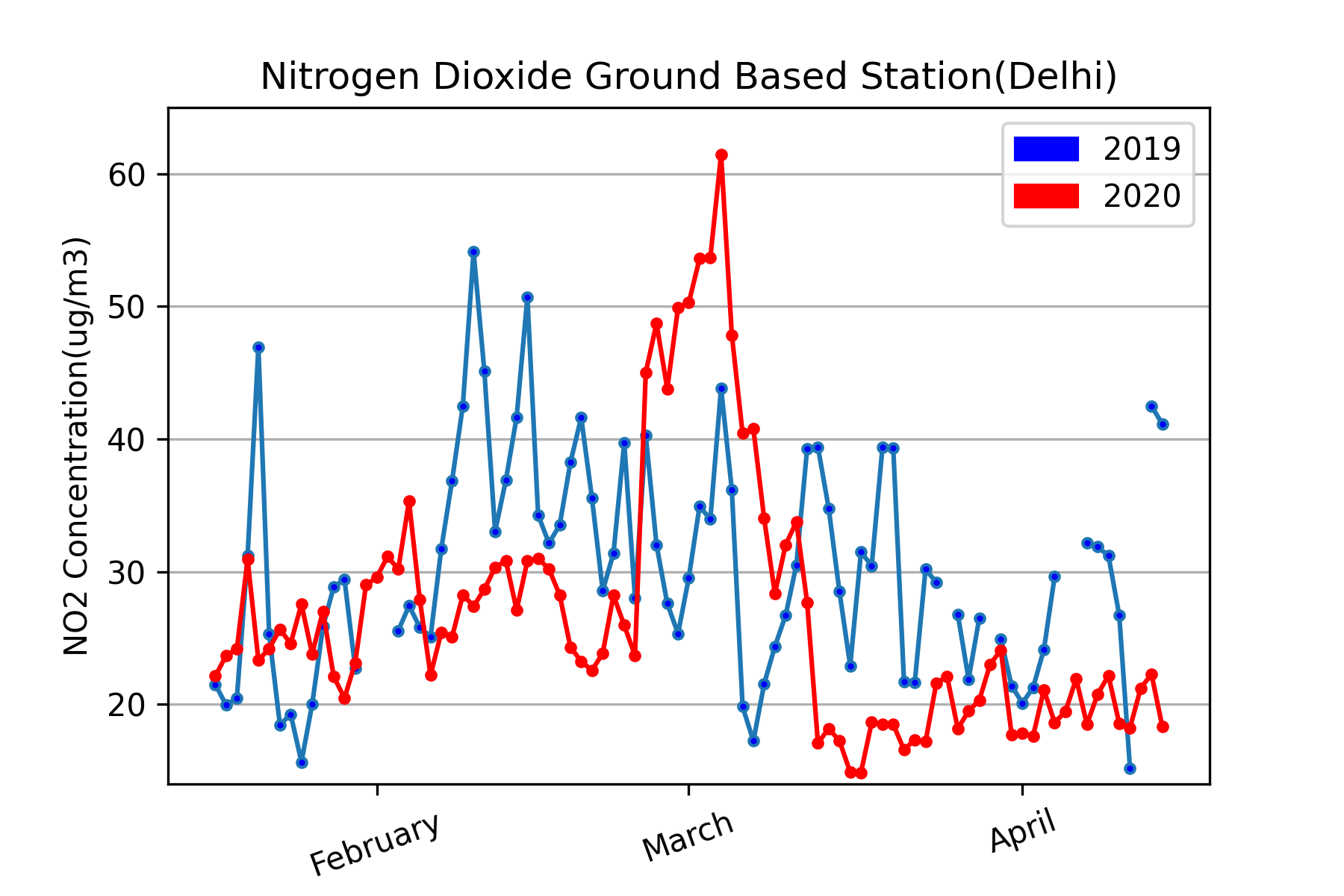}}
\subfloat[Sulphur Dioxide]{\includegraphics[height=0.35\textwidth]{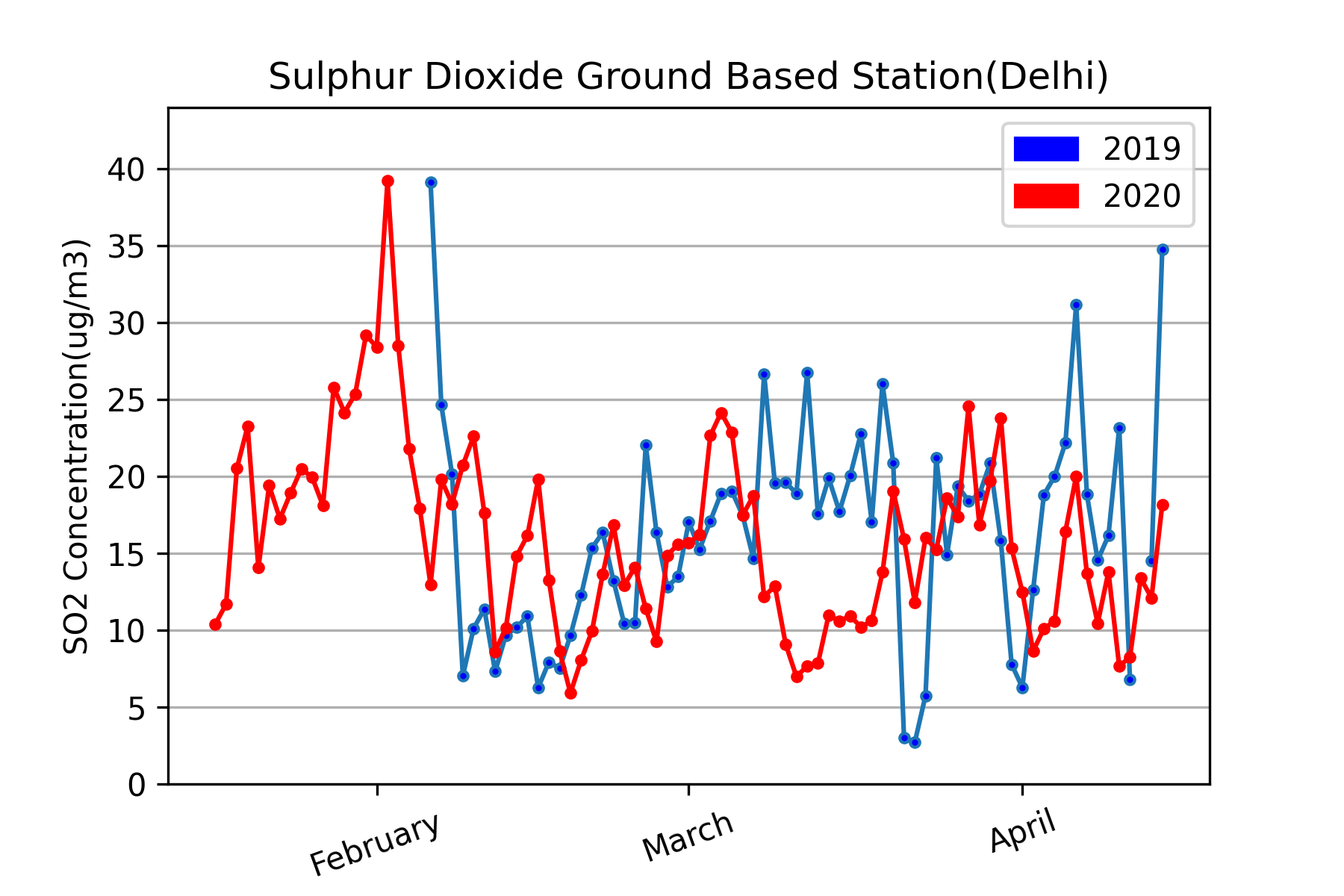}}\\
\subfloat[Carbon Monoxide]{\includegraphics[height=0.35\textwidth]{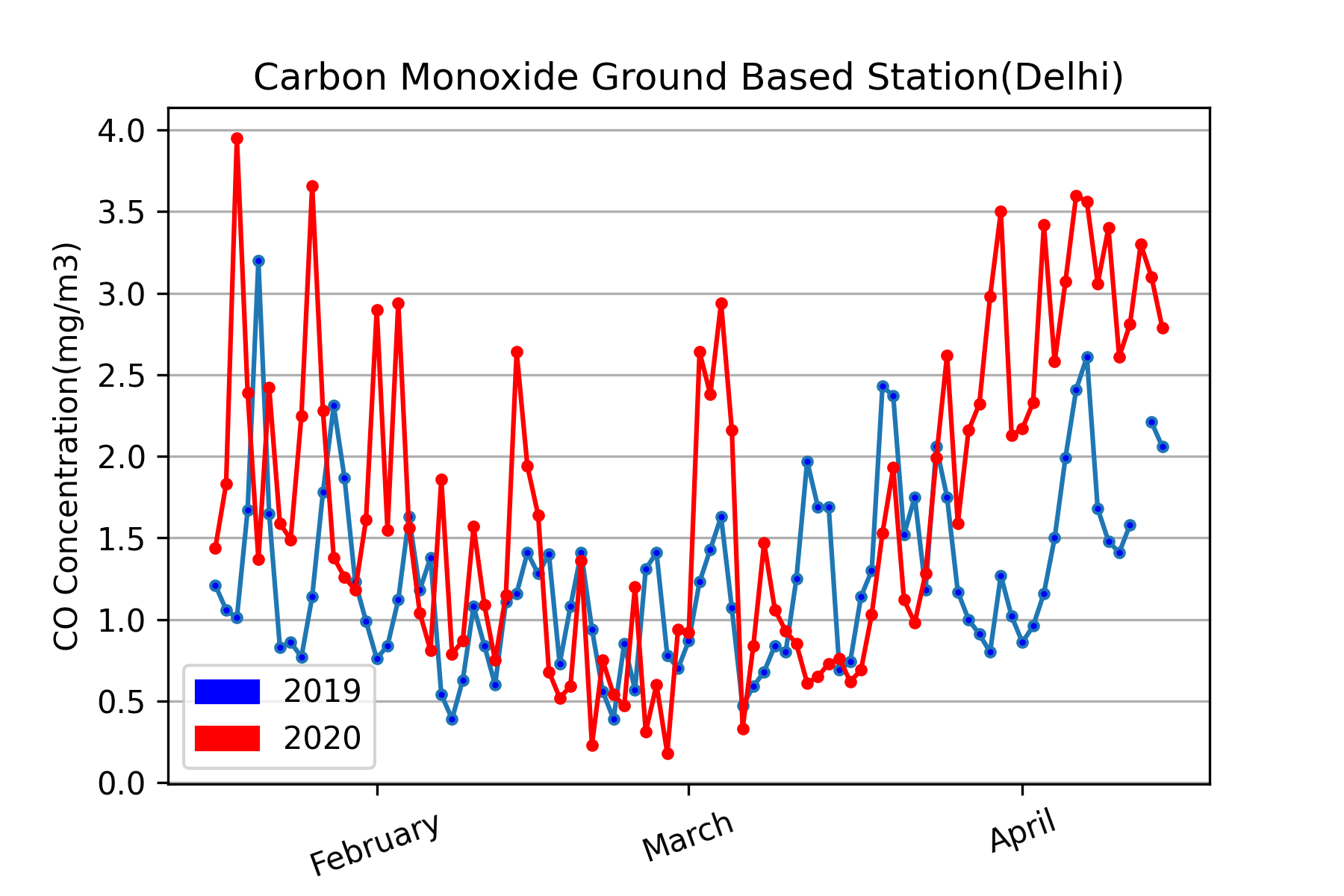}}
\subfloat[Ozone]{\includegraphics[height=0.35\textwidth]{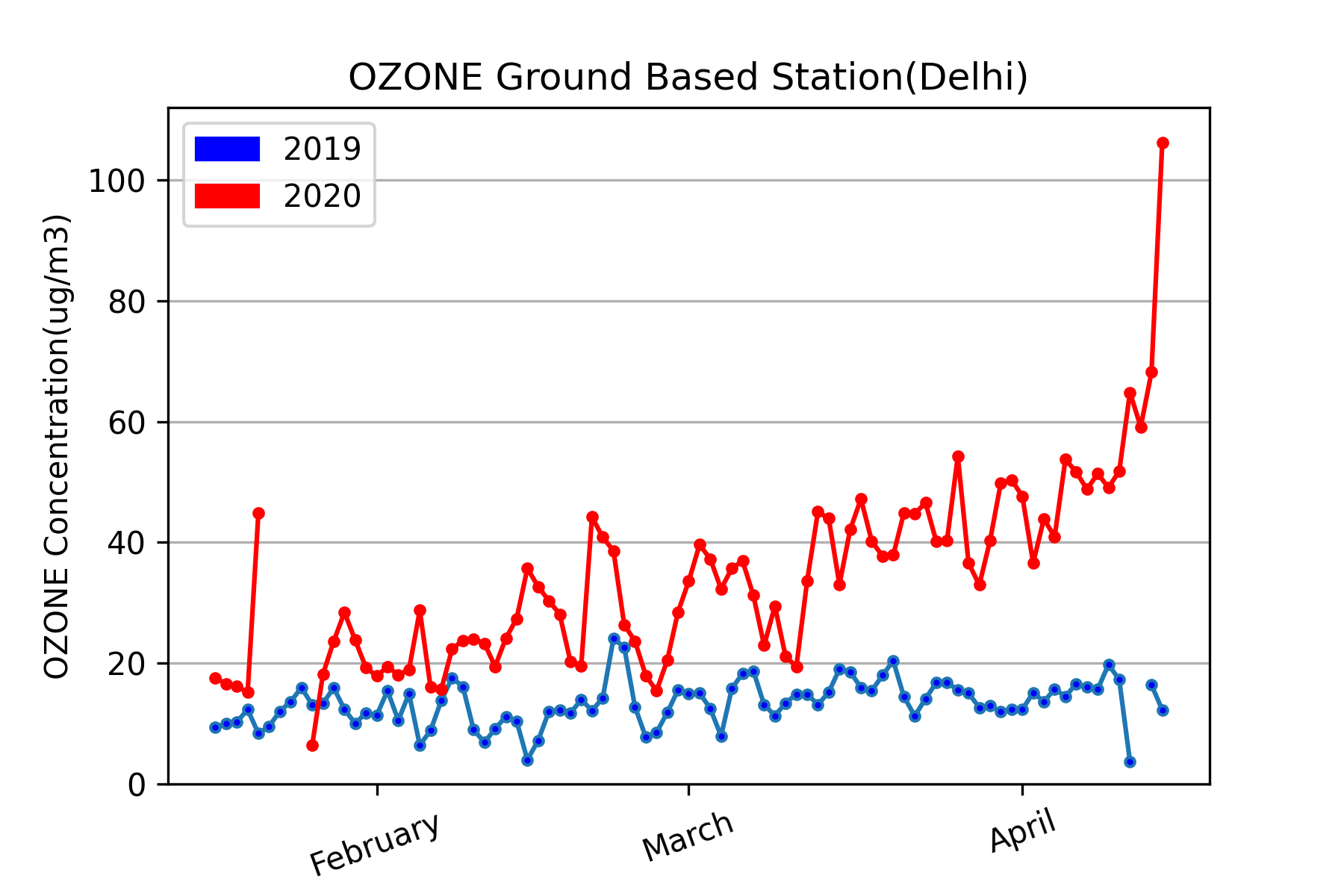}}
\caption{Trend of various atmospheric pollutants obtained from ground-based observations across the lockdown period. We observed that (a) concentration of NO$_2$ had a peak of $61.47$ $\mu$g/m$^3$ on 20-Mar-2020, and then continued to drop, (b) concentration of SO$_2$ showed a drop during initial days of lockdown, (c) concentration of CO showed a drop from $2.94$ mg/m$^3$ on 20-Mar-2020 to $0.33$ mg/m$^3$ on 22-Mar-2020 (day of \textit{janta curfew}) recording a drop of approximately 88\%, and (d) the concentration of O$_3$ was at its minimum concentration of $15.41$ $\mu$g/m$^3$ on 14-Mar-2020 and peaked to $45.06$ $\mu$g/m$^3$ on 29-Mar-2020. The concentration kept increasing subsequently.}
\label{fig:ground}
\end{figure*}

The particulate matter data collected from these ground-based
monitoring stations indicated an improvement in the PM
concentration~\cite{explo}. The PM2.5 concentration in Delhi
before the lockdown  was in the range of
$60\mbox{ }\mu$g/m$^3$ to $125\mbox{ }\mu$g/m$^3$. However,
after the lockdown period, the concentration clearly dropped by
about 50\% to the range of $25\mbox{ }\mu$g/m$^3$ to
$60\mbox{ }\mu$g/m$^3$. A similar drop was visible in the
concentration of PM10, which was recorded to be
$175\mbox{ }\mu$g/m$^3$ on 20-Mar-2020. This reduced to
$100\mbox{ }\mu$g/m$^3$ on 22-Mar-2020 and
$60\mbox{ }\mu$g/m$^3$ on 26-Mar-2020, following the similar
decreasing trend  for the period of
lockdown~\cite{mahato2020effect}.

In addition to this particulate matter concentration, we observed similar trend in the concentration of harmful gases in the atmosphere, which is in line with our analysis on Sentinel-5P satellite data.   In  Fig.~\ref{fig:ground}, we observed that the concentration of NO$_2$ significantly changed throughout the month of Mar-2020. The NO$_2$ concentration in the air had a peak of $61.47\mbox{ }\mu$g/m$^3$ on 20-Mar-2020 (just two days before \textit{janta curfew}), which dropped to $40.43\mbox{ }\mu$g/m$^3$ on 22-Mar-2020 (date of \textit{janta curfew}) and continued to drop to about $17.25\mbox{ }\mu$g/m$^3$ on 31-Mar-2020.  Consequently, the average concentration of NO$_2$ saw a significant drop from $61.47\mbox{ }\mu$g/m$^3$ to $17.25\mbox{ }\mu$g/m$^3$ during the month of Mar-2020. We observed similar trend in the reduction of NO$_2$ concentration when compared between the same days of 2019 and 2020.

Similarly, Sulfur Dioxide (SO$_2$) concentration in the air had a peak of $24.16\mbox{ }\mu$g/m$^3$ on 20-Mar-2020 (just two days before \textit{janta curfew}), which dropped to $17.48\mbox{ }\mu$g/m$^3$ on 22-Mar-2020 (day of \textit{janta curfew}) and continued to drop to $7.67\mbox{ }\mu$g/m$^3$ on 28-Mar-2020. The ground based data showed an abnormal increase in Sulfur Dioxide concentration for the first half of April which subsequently dropped noticeably in later April. Carbon Monoxide (CO) is generally produced from incomplete combustion of fuels containing carbon. Ground based collected data showed that the concentration of CO dropped to 2.94\mbox{ }mg/m$^3$ on 20-Mar-2020, and further dropped to 0.33\mbox{ }mg/m$^3$ on 22-Mar-2020 (day of \textit{janta curfew}) indicating a drop of approximately 88\%. This decrease continued till the end of Mar-2020 and first week of Apr-2020.

We also obtained the daily average ozone concentration in the atmosphere from the ground-based pollutant data. We observed from the ground-based data that O$_3$ concentration kept on rising post lockdown. The minimum concentration for month of Mar-2020 was $15.41$ $\mu$g/m$^3$ on 14-Mar-2020 and had a peak of $45.06$ $\mu$g/m$^3$ on 29-Mar-2020 which was one of the initial day of lockdown. These values kept on increasing and reached maximum value of $106.23$ $\mu$g/m$^3$(for April) on 30-Apr-2020.  This trend is in agreement with the satellite-based O$_3$ concentration analysis.

Our study indicated that the restrictions put in place because of COVID-19 had a clear positive impact on the pollutant concentration in the atmosphere. Owing to these restrictions, there was a reduced human movement, reduced vehicular emissions, and less industrial activities. Such studies can pave the path for the government (both at national and state level) to enact laws in dramatically reducing emission of toxic air pollutants, and provide healthcare protections to the citizens. These steps can include reducing toxic air pollution from the industrial sources, enacting stringent standards for vehicles and engines, and educating the general public via community-driven voluntary programmes. Such laws were enacted by the Government of India via the Motor Vehicles Act, 1988, that required petrol driven vehicles to require PUC (Pollution Under Control) certificate for low emission of carbon monoxide in the atmosphere. Therefore, our study will greatly assist the environmentalists in better understanding the atmosphere.

\section{Conclusion \& Future Works}

\label{sec:conc}
Delhi being one of the most polluted cities observed a drastic and noticeable drop in harmful gases like Nitrogen Dioxide, Carbon Monoxide, and Sulfur Dioxide. An interesting phenomena noticed in the atmosphere of Delhi was the increase in ozone concentration as the lockdown was imposed which was also linked to the improvement in the atmospheric conditions comparatively.  Obtaining satellite data from Sentinel-5P depicted the scenario from a different perspective, where ground based station data verified the obtained results exactly. 
Remote location with lack of ground based monitoring stations will benefit from such remote sensing techniques for atmospheric monitoring without any additional setup cost.  The COVID-19 lockdown brought a rare opportunity for researchers to understand the environment much better and thereby attempt to influence the policy makers to implement stricter environment laws. 

Our future work include the short-term forecasting of pollutant
data from historical data. We intend to use the popular
LSTM-based models~\cite{pathan2021efficient,jain2020forecasting}
to learn the underlying pattern of such time-series pollutant
data. Furthermore, we also plan to mine COVID-19 related texts in
different languages~\cite{berjon2021analysis}, and provide an
universal knowledge bank. We also intend to understand the impact
of other real-life affecting factors, including population size,
urban/rural divide, level of industrialization, income, and
pulmonary diseases, on the air quality of a megacity. Such
studies will enable us in proposing relevant recommendations to
city planners and environmental agencies.

\section*{Acknowledgments}
This research was conducted with the financial support of Science Foundation Ireland under Grant Agreement No.\ 13/RC/2106\_P2 at the ADAPT SFI Research Centre at University College Dublin. ADAPT, the SFI Research Centre for AI-Driven Digital Content Technology, is funded by Science Foundation Ireland through the SFI Research Centres Programme.

\balance


\begin{thebibliography}{10}
\expandafter\ifx\csname url\endcsname\relax
  \def\url#1{\texttt{#1}}\fi
\expandafter\ifx\csname urlprefix\endcsname\relax\def\urlprefix{URL }\fi
\expandafter\ifx\csname href\endcsname\relax
  \def\href#1#2{#2} \def\path#1{#1}\fi

\bibitem{agrawal2021investigation}
P.~Agrawal, G.~Kaur, S.~S. Kolekar, Investigation on biomedical waste
  management of hospitals using cohort intelligence algorithm, Soft Computing
  Letters 3 (2021) 100008.

\bibitem{guegana2018sustainable}
J.-F. Gu{\'e}gana, G.~Suz{\'a}n, S.~Kati-Coulibaly, D.~N. Bonpamgue, J.-P.
  Moatti, Sustainable development goal\# 3,“health and well-being”, and the
  need for more integrative thinking, Veterinaria M{\'e}xico 5~(2) (2018)
  1--18.

\bibitem{vaidya2020sdg}
H.~Vaidya, T.~Chatterji, {SDG} 11 sustainable cities and communities, in:
  Actioning the Global Goals for Local Impact, Springer, 2020, pp. 173--185.

\bibitem{singh2021parallel}
K.~R. Singh, K.~Neethu, K.~Madhurekaa, A.~Harita, P.~Mohan, Parallel {SVM}
  model for forest fire prediction, Soft Computing Letters (2021) 100014.

\bibitem{jia2020insignificant}
C.~Jia, X.~Fu, D.~Bartelli, L.~Smith, Insignificant impact of the
  “stay-at-home” order on ambient air quality in the memphis metropolitan
  area, usa, Atmosphere 11~(6) (2020) 630.

\bibitem{zangari2020air}
S.~Zangari, D.~T. Hill, A.~T. Charette, J.~E. Mirowsky, Air quality changes in
  {New} {York} city during the {COVID-19} pandemic, Science of the Total
  Environment 742 (2020) 140496.

\bibitem{bourdrel2021impact}
T.~Bourdrel, I.~Annesi-Maesano, B.~Alahmad, C.~N. Maesano, M.-A. Bind, The
  impact of outdoor air pollution on {COVID}-19: a review of evidence from in
  vitro, animal, and human studies, European Respiratory Review 30~(159)
  (2021).

\bibitem{ali2020effects}
N.~Ali, F.~Islam, The effects of air pollution on {COVID}-19 infection and
  mortality—a review on recent evidence, Frontiers in Public Health 8 (2020).

\bibitem{nigam2021positive}
R.~Nigam, K.~Pandya, A.~J. Luis, R.~Sengupta, M.~Kotha, Positive effects of
  {COVID}-19 lockdown on air quality of industrial cities ({Ankleshwar} and
  {Vapi}) of {Western} {India}, Scientific Reports 11~(1) (2021) 1--12.

\bibitem{karaer2020analyzing}
A.~Karaer, N.~Balafkan, M.~Gazzea, R.~Arghandeh, E.~E. Ozguven, Analyzing
  COVID-19 impacts on vehicle travels and daily nitrogen dioxide (no2) levels
  among florida counties, Energies 13~(22) (2020) 6044.

\bibitem{kaloni2021impact}
D.~Kaloni, Y.~H. Lee, S.~Dev, Impact of {COVID}19-induced lockdown on air
  quality in {Ireland}, in: Proc. International Geoscience and Remote Sensing
  Symposium (IGARSS), 2021.

\bibitem{franch2021review}
I.~Franch-Pardo, M.~R. Desjardins, I.~Barea-Navarro, A.~Cerd{\`a}, A review of
  {GIS} methodologies to analyze the dynamics of {COVID}-19 in the second half
  of 2020, Transactions in GIS (2021).

\bibitem{doolette2020zinc}
C.~Doolette, T.~Read, N.~Howell, T.~Cresswell, E.~Lombi, Zinc from
  foliar-applied nanoparticle fertiliser is translocated to wheat grain: a 65zn
  radiolabelled translocation study comparing conventional and novel foliar
  fertilisers, Science of The Total Environment 749 (2020) 142369.

\bibitem{alparslan2021analyzing}
B.~Alparslan, M.~Jain, J.~Wu, S.~Dev, Analyzing air pollutant concentrations in
  New Delhi, India, in: Proc. Progress in Electromagnetics Research
  Symposium-Fall (PIERS-FALL), 2021.

\bibitem{cngreport}
The impact of {Delhi}'s {CNG} program on air quality,
  \url{https://media.rff.org/documents/RFF-DP-07-06.pdf}.

\bibitem{mathur2019impact}
S.~K. Mathur, P.~Murali~Prasad, P.~Kulshreshtha, S.~Khorana, M.~Chauhan, The
  impact of odd-even transportation policy and other factors on pollution in
  {Delhi}: A spatial and {RDD} analysis, Tech. rep., ARTNeT Working Paper
  Series (2019).

\bibitem{mohan2007analysis}
M.~Mohan, A.~Kandya, An analysis of the annual and seasonal trends of air
  quality index of {Delhi}, Environmental monitoring and assessment 131~(1)
  (2007) 267--277.

\bibitem{explo}
N.~Khan, \href{http://pubs.sciepub.com/}{An exploratory study of impact of
  lockdown on the air quality of {Delhi}}, Applied Ecology and Environmental
  Sciences 8~(5) (2020) 261--268.
\newline\urlprefix\url{http://pubs.sciepub.com/}

\bibitem{guanter2015potential}
L.~Guanter, I.~Aben, P.~Tol, J.~Krijger, A.~Hollstein, P.~K{\"o}hler, A.~Damm,
  J.~Joiner, C.~Frankenberg, J.~Landgraf, Potential of the {TROPO}spheric
  {M}onitoring {I}nstrument ({TROPOMI}) onboard the {Sentinel-5 Precursor} for
  the monitoring of terrestrial chlorophyll fluorescence, Atmospheric
  Measurement Techniques 8~(3) (2015) 1337--1352.

\bibitem{tropomidetails}
Tropomi instrument details, \url{http://www.tropomi.eu/}.

\bibitem{sun2015slope}
Q.~Sun, L.~Zhang, X.~Ding, J.~Hu, Z.~Li, J.~Zhu, Slope deformation prior to
  {Zhouqu}, {China} landslide from {InSAR} time series analysis, Remote Sensing
  of Environment 156 (2015) 45--57.

\bibitem{das2021estimating}
B.~P. Das, M.~S. Pathan, Y.~H. Lee, S.~Dev, Estimating ground-level nitrogen
  dioxide concentration from satellite data, in: Proc. Progress in
  Electromagnetics Research Symposium-Fall (PIERS-FALL), 2021.

\bibitem{akrami2021graph}
N.~Akrami, K.~Ziarati, S.~Dev, Graph-based local climate classification in
  {Iran}, International Journal of Climatology (2021).

\bibitem{manandhar2019data}
S.~Manandhar, S.~Dev, Y.~H. Lee, Y.~S. Meng, S.~Winkler, A data-driven approach
  for accurate rainfall prediction, IEEE Transactions on Geoscience and Remote
  Sensing 57~(11) (2019) 9323--9331.

\bibitem{dev2019estimating}
S.~Dev, F.~M. Savoy, Y.~H. Lee, S.~Winkler, Estimating solar irradiance using
  sky imagers, Atmospheric Measurement Techniques 12~(10) (2019) 5417--5429.

\bibitem{dev2017color}
S.~Dev, Y.~H. Lee, S.~Winkler, Color-based segmentation of sky/cloud images
  from ground-based cameras, IEEE Journal of Selected Topics in Applied Earth
  Observations and Remote Sensing 10~(1) (2017) 231--242.

\bibitem{danesi2021monitoring}
N.~Danesi, M.~Jain, Y.~H. Lee, S.~Dev, Monitoring atmospheric pollutants from
  ground-based observations, in: Proc. IEEE AP-S Symposium and USNC-URSI Radio
  Science Meeting, IEEE, 2021.

\bibitem{srivastava202021}
S.~Srivastava, A.~Kumar, K.~Bauddh, A.~S. Gautam, S.~Kumar, 21-day lockdown in
  {India} dramatically reduced air pollution indices in {Lucknow} and {New}
  {Delhi}, {India}, Bulletin of Environmental Contamination and Toxicology
  (2020) 1.

\bibitem{nagappa2020now}
B.~Nagappa, M.~Sakthivel, Y.~Marimuthu, A.~Rastogi, A.~Ramalingam, S.~K. Sarin,
  Now casting and forecasting of {COVID}-19 outbreak in the national capital
  region of {Delhi}, medRxiv (2020).

\bibitem{NO2Sent1}
Copernicus sentinel data processed by {ESA}, {Koninklijk} {Nederlands}
  {Meteorologisch} {Instituut} ({KNMI}) (2018), {Sentinel}-{5P} {TROPOMI}
  {Tropospheric} {NO2} 1-orbit l2 7km x 3.5km, {G}reenbelt, {MD}, {USA},
  {G}oddard {E}arth {S}ciences {D}ata and {I}nformation {S}ervices {C}enter
  ({GES DISC}), \url{10.5270/S5P-s4ljg54}, accessed: 2020-06-20.

\bibitem{SO2Sent1}
Copernicus {S}entinel data processed by {ESA}, {G}erman {A}erospace {C}enter
  ({DLR}) (2019), {S}entinel-5{P} {TROPOMI} {S}ulphur {D}ioxide {SO2} 1-{O}rbit
  {L}2 5.5km x 3.5km, {G}reenbelt, {MD}, {USA}, {G}oddard {E}arth {S}ciences
  {D}ata and {I}nformation {S}ervices {C}enter ({GES DISC}),
  \url{10.5270/S5P-yr8kdpp}, accessed: 2020-06-20.

\bibitem{So2ProductManual}
S5p mission performance centre sulphur dioxide,
  \url{https://sentinels.copernicus.eu/documents/247904/3541451/Sentinel-5P-Sulphur-Dioxide-Readme.pdf}.

\bibitem{COSent}
Copernicus {S}entinel data processed by {ESA}, {K}oninklijk {N}ederlands
  {M}eteorologisch {I}nstituut ({KNMI})/{N}etherlands {I}nstitute for {S}pace
  {R}esearch ({SRON}) (2019), {S}entinel-{5P} {TROPOMI} {C}arbon {M}onoxide
  {CO} {C}olumn 1-{O}rbit l2 7km x 7km, {G}reenbelt, {MD}, {USA}, {G}oddard
  {E}arth {S}ciences {D}ata and {I}nformation {S}ervices {C}enter ({GES DISC}),
  \url{10.5270/S5P-1hkp7rp}, accessed: 2020-06-20.

\bibitem{airqual2020}
C.~India, Air quality analysis during summer lockdown: Some highlights,
  Discussion paper (2020).

\bibitem{OZONESent1}
Copernicus {S}entinel data processed by {ESA}, {G}erman {A}erospace {C}enter
  ({DLR}) (2019), {S}entinel-5{P} {TROPOMI} {T}otal {O}zone {C}olumn 1-{O}rbit
  l2 5.5km x 3.5km, {G}reenbelt, {MD}, {USA}, {G}oddard {E}arth {S}ciences
  {D}ata and {I}nformation {S}ervices {C}enter ({GES DISC}),
  \url{10.5270/S5P-fqouvyz}, accessed: 2020-06-20.

\bibitem{mahato2020effect}
S.~Mahato, S.~Pal, K.~G. Ghosh, Effect of lockdown amid {COVID-19} pandemic on
  air quality of the megacity {Delhi}, {India}, Science of the Total
  Environment (2020) 139086.

\bibitem{pathan2021efficient}
M.~S. Pathan, M.~Jain, Y.~H. Lee, T.~AlSkaif, S.~Dev, Efficient forecasting of
  precipitation using {LSTM}, in: Proc. Progress in Electromagnetics Research
  Symposium-Fall (PIERS-FALL), 2021.
  
\bibitem{danesi2021predicting}
N. Danesi, M. Jain, Y. H. Lee, S. Dev, Predicting Ground-based PM$_{2.5}$ Concentration in Queensland, Australia, in: Proc. Progress in Electromagnetics Research
  Symposium-Fall (PIERS-FALL), 2021.

\bibitem{jain2020forecasting}
M.~Jain, S.~Manandhar, Y.~H. Lee, S.~Winkler, S.~Dev, Forecasting precipitable
  water vapor using {LSTM}s, in: Proc. IEEE AP-S Symposium and USNC-URSI Radio
  Science Meeting, IEEE, 2020.

\bibitem{berjon2021analysis}
P.~Berjon, A.~Nag, S.~Dev, Analysis of {French} phonetic idiosyncrasies for
  accent recognition, Soft Computing Letters (2021) 100018.

\end{thebibliography}
\end{document}